# COMMUNICATION

# Spin-momentum locking induced non-local voltage in topological insulator nanowire

Jen-Ru Chen,[a] Pok Lam Tse,[b] Ilya N. Krivorotov,[a] and Jia G Lu *[b]

**The momentum and spin of charge carriers in the topological insulators are constrained to be perpendicular to each other due to the strong spin-orbit coupling. We have investigated this unique spin-momentum locking property in $Sb_2Te_3$ topological insulator nanowires by injecting spin-polarized electrons through magnetic tunnel junction electrodes. Non-local voltage measurements exhibit a symmetry with respect to the magnetic field applied perpendicular to the nanowire channel, which is remarkably different from that of a non-local measurement in a channel that lacks spin-momentum locking. In stark contrast to conventional non-local spin valves, simultaneous reversal of magnetic moments of all magnetic contacts to the $Sb_2Te_3$ nanowire alters the non-local voltage. This unusual symmetry is a clear signature of the spin-momentum locking in the $Sb_2Te_3$ nanowire surface states.**

## 1. Introduction

Topological insulator (TI) is a crystal, in which strong spin-orbit interaction due to heavy elements, such as Bi, Se, Sb, Te, Sn and Pb, gives rise to electronic band inversion and a topologically protected surface state.[1] It is an insulator in its bulk, while the gapless surface state with linear energy-momentum dispersion relation gives rise to surface conduction.[2,3] Electronic transport in TI surface states has another remarkable characteristic – electrons with opposite spins propagate in opposite directions because spin is locked at right angle to the momentum.[4,5] This spin-momentum locking leads to perfect spin polarization of surface currents, which yields high efficiency of spin torques generated by TIs.[6-10] The TI surface states obey time-reversal symmetry, and since backscattering in the surface state requires a spin flip, it is forbidden in spin-conserving scattering processes.

Quasi-1D TI nanowires is an attractive system for studies of non-trivial topological surface states. In short, the advantages of the quasi-1D system are manifold, including the suppression of the bulk conductivity due to the high surface-to-bulk ratio,[11] discrete 1D sub-bands for control of the transmission modes, and Fermi level can be effectively tuned between *p* and *n* types by chemical doping [12-15] or band structure engineering.[16-19] Moreover, magnetic field provides additional tunability of the surface state in nanowires.[20-22] The nanowire cross section small enough to open a sizeable gap for the surface states, yet large enough that a moderate magnetic field parallel to the wire can thread a half flux quantum *h/2e* to restore the gapless 1D mode.[23-25]

$Sb_2Te_3$ is a topological insulator, which has a bulk band gap of 0.28 eV and a simple surface states consisting of a single Dirac cone in the band gap. The pristine crystalline structure of $Sb_2Te_3$ is hexagonal, and the primitive cell is rhombohedral ($R\bar{3}m$). Our previous studies on the nanowires have revealed the single crystalline structure with repeating quintuple layers of (Te-Sb-Te-Sb-Te).[25-27] We have also performed low temperature magnetoresistance measurements and angle resolved photoemission spectroscopy on these nanowires synthesized by the same setup as presented in this work. The periodic Aharonov-Bohm type oscillations observed manifest the transport in topologically protected surface states in the *p*-type $Sb_2Te_3$ nanowires, with a Fermi level positioned approximately 40 meV below the $\Gamma$-point.[25, 26]

## 2. Experimental

### 2.1 Synthesis and structural characterization

$Sb_2Te_3$ nanowires were synthesized by low pressure catalytic chemical vapor deposition via vapor-liquid-solid (VLS) growth mechanism.[28] Two source materials: 0.6 g antimony powder and 1.0 g tellurium powder, were placed upstream at the centre of the heating zone in a quartz tube, and Au catalyst deposited

[a.] *Department of Physics and Astronomy, University of California, Irvine, California 92697, USA*
[b.] *Department of Physics and Astronomy and Department of Electrophysics, University of Southern California, Los Angeles, CA 90089, USA. Email: jialu@usc.edu*

**COMMUNICATION**

Si/SiO$_2$ substrate was placed downstream. Argon carrier gas was supplied at a flow rate of 80 standard cubic centimeter per minute. The growth process at 430 °C lasted about 6 hours.

Transmission electron microscopy (TEM) image of the cross-section of a pristine Sb$_2$Te$_3$ nanowire, as shown in Fig. 1a, indicates quintuple layer (Te-Sb-Te-Sb-Te) stacking with an interlayer van der Waals gap of 0.309 nm. X-ray diffraction (XRD) spectrum of the nanowire is plotted in Fig. 1b, which verifies that the nanowire has rhombohedral ($R\bar{3}m$) crystal structure (JCPDS PDF#15-0874). Fig. 1c depicts a scanning electron microscopy (SEM) image of a single Sb$_2$Te$_3$ with a length of approximately 5 µm and width of 200 nm. Fig. 1d displays the enlarged view of the nanowire tip region, showing the Au catalyst capped nanoparticle, confirming the VLS tip-growth mechanism. Energy-dispersive X-ray Spectroscopy (EDS) results indicate that the atomic ratio of Sb:Te is 2:3 and the mapping of Sb and Te elemental signals along a single nanowire is illustrated in Fig. 1e and Fig. 1f, demonstrating the uniform distribution of Sb and Te along the nanowire.

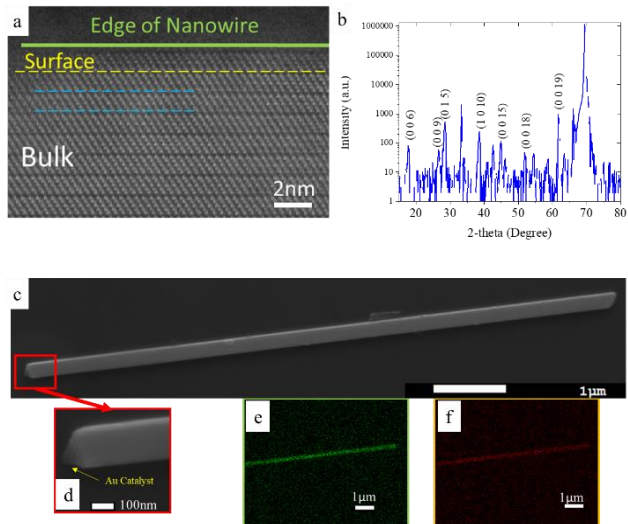

**Fig. 1** TEM and SEM images of Sb$_2$Te$_3$ nanowire (a) High Resolution TEM image of the cross-section of a Sb$_2$Te$_3$ nanowire showing quintuple layers (Te-Sb-Te-Sb-Te) stacking and the van der Waals gaps in between quintuple layers (blue dashed lines). (b) XRD spectrum of Sb$_2$Te$_3$ with peaks representing crystal planes of Rhombohedral ($R\bar{3}m$) crystal structure (JCPDS PDF#15-0874) (c) SEM image of a Sb$_2$Te$_3$ nanowire with length ~5 µm. (d) Enlarged view of the nanowire tip showing the Au catalyst, indicating VLS growth. (e) & (f) respective EDX mapping of nanowire's Te signal (green) and Sb signal (red), showing that Sb and Te are uniformly distributed along the nanowire.

**2.2 Device fabrication**

In order to probe spin dependent transport through topological surface states, we have examined the non-local voltage signal in a TI Sb$_2$Te$_3$ channel, as illustrated in Fig. 2. To carry out such measurements, a nanodevice is fabricated, consisting of a Sb$_2$Te$_3$ nanowire with two ferromagnetic leads (labeled E2 and E3), and two non-magnetic leads (labeled E1 and E4) attached to the Sb$_2$Te$_3$ wire. The Sb$_2$Te$_3$ nanowire has rectangular cross-section[25] (data not shown), with dimension ~5 µm long, ~100 nm wide and ~50 nm thick. The edge-to-edge separation between the two ferromagnetic leads (E2 and E3) is 0.5 µm. In the first step of the device fabrication process, the Sb$_2$Te$_3$ nanowires are dispersed onto a thermally oxidized Si substrate.

Individual nanowires are then coordinated with respect to the alignment marks on the Si wafer via SEM imaging. Then two steps of aligned e-beam lithography are utilized to pattern the non-magnetic Nb (5 nm)/Au (35 nm) outer leads that form ohmic contacts to the Sb$_2$Te$_3$ nanowire and two magnetic tunnel junction inner leads: AlO$_x$ (1.5 nm)/Ni$_{80}$Fe$_{20}$ (20 nm)/Cu (4 nm)/Co (5 nm)/CoO (2 nm). The tunnel junction leads consist of several layers. Magnetization direction of the Permalloy (Ni$_{80}$Fe$_{20}$=Py) free layer determines the magnitude and direction of spin current polarization injected into the Sb$_2$Te$_3$ nanowire via the AlO$_x$ tunnel barrier. Py is a magnetically soft material and its magnetization can be easily switched by a low external magnetic field. Magnetization direction of the Co pined layer is strongly pinned by exchange bias from antiferromagnetic CoO layer.[29] The pinned Co layer serves as a reference that allows unambiguous determination of the Py layer magnetization direction via measuring the current-in-plane giant magneto-resistance of the lead itself.

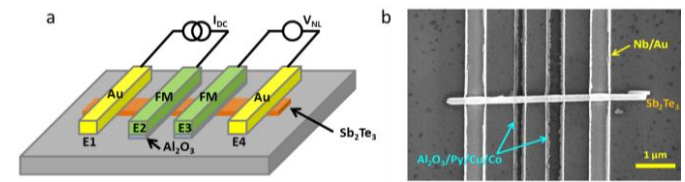

**Fig. 2** (a) Schematic of the nonlocal spin valve device based on a Sb$_2$Te$_3$ nanowire channel. Two inner AlO$_x$/Py/Cu/Co/CoO ferromagnetic electrodes (FM) form magnetic tunnel junction nanocontacts to the TI nanowire channel. Two outer non-magnetic Nb/Au electrodes make ohmic contact to the nanowire. The non-local voltage $V_{NL}$ between the contacts labeled E3 and E4 is generated in response to a direct current $I_{DC}$ applied between the contacts E1 and E2. (b) SEM image of a device.

**2.3 Magneto-transport measurements**

We have proven earlier in our samples, from the comparison of 2-probe and 4-probe measurements, that the contact between Au electrode to the nanowire is of ohmic nature with negligible contact resistance. Thus, the resistance measurement presented here is carried out by 2-probe. The resistance measured at $T$ = 4.2 K between the Au contacts as a function of magnetic field $H$ applied parallel to the Sb$_2$Te$_3$ nanowire axis, as exhibited in Fig. 3a, indicates positive magneto-resistance (defined as [$R(H)$-$R(0)$]/$R(0)$, where $R(H)$ is the two-point resistance measured at magnetic field $H$). This positive magneto-resistance originates from weak anti-localization of carriers in the Sb$_2$Te$_3$ nanowire induced by spin-orbit interaction.[30] This weak anti-localization signal demonstrates strong impact of the spin orbit interaction on transport in our Sb$_2$Te$_3$ nanowire system. Fig. 3b shows the temperature dependence of the magneto-resistance. The magnitude of the magneto-resistance decreases with increasing temperature by more than a factor of 5 between 4.2 K and 120 K. This decrease of the weak anti-localization magneto-resistance arises from temperature-induced decoherence of conduction charges.

**2.4 Non-local voltage measurements**

We have measured the non-local voltage in the multi-contacted Sb$_2$Te$_3$ nanowire device at T = 4.2 K, as illustrated in Fig. 4. In these measurements, a direct electric current $I_{DC}$ is applied between two left leads (E1 and E2), and the non-local voltage $V_{NL}$ is measured between the pair of right leads (E3 and E4).[31-33] Injection of a spin-polarized electric current through the tunnel junction contact E2 into the nanowire gives rise to spin current in the TI surface states. Owing to spin-momentum locking, this spin current flows either to the left or to the right of the injector E2 (towards or away from the detector E3), depending on the polarity of $I_{DC}$ and the direction of the Py injector electrode magnetization $M_{Py}$. Reversal of $I_{DC}$ polarity or reversal of $M_{Py}$ inverts the direction of the spin current in the nanowire. Given the dependence of spin current on $I_{DC}$ polarity and $M_{Py}$, spin accumulations also shift either towards or away from the detector E3, depending on the $I_{DC}$ polarity and the $M_{Py}$ direction. This spin accumulation at the detector electrode E3 modifies electric charge accumulation at E3, and in consequence, changes $V_{NL}$ measured at E3. This spin current driven charge accumulation at the detector electrode explains the symmetry with respect to the magnetic field and inversion upon reversal of the bias current or electrode magnetization in the non-local voltage signals revealed in our TI nanochannel.

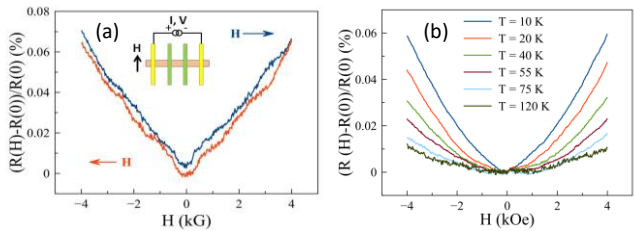

**Fig. 3** (a) Positive magneto-resistance of the $Sb_2Te_3$ nanowire measured between two non-magnetic ohmic leads at T = 4.2 K arises from weak anti-localization of carriers in the nanowire induced by spin-orbit interaction. Inset illustrates the applied magnetic field direction as well as the two-probe measurement circuitry employed. (b) Temperature dependence of the magneto-resistance.

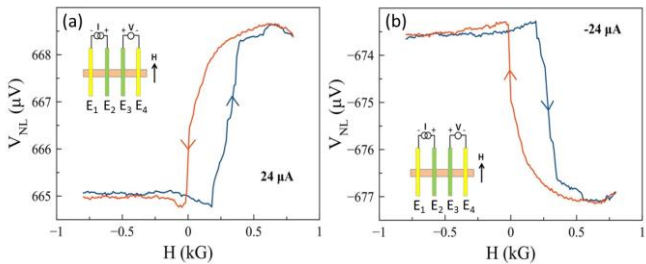

**Fig. 4** Non-local voltage $V_{NL}$ in the $Sb_2Te_3$ nanowire device measured between leads E3 and E4 for (a) positive and (b) negative current bias $I_{DC}$ =± 24 μA applied between the left pair of leads (E1 and E2) as illustrated in the insets. Magnetic field is applied in the plane of the sample parallel to the magnetic electrode wires (perpendicular to the $Sb_2Te_3$ nanowire). The Py layer magnetic moments in the two magnetic layers switch at nearly identical magnetic fields.

Let us examine the cases in more details: four different spin valve states are displayed in Fig. 5, which differ by the current $I_{DC}$ polarity applied to the injector and the direction of saturated Py layer magnetic moments $M_{Py}$. Tunnelling is governed by the density of states of electrons at the Fermi level. And for permalloy, the Fermi level is situated at the minority band.[34] Therefore, when the injected electrons tunnelling into (or out of) the nanowire channel with polarization down (or up), as depicted in Fig. 5b (or 5e), the excess spin down electrons in the channel, in either case, flow in the direction toward the detector owing to the spin ($s$) – momentum ($k$) locking, yielding high electric charge accumulation under the detector Py electrode. Likewise, if the injected electrons into (or out of) the nanochannel have polarizations up (or down), as depicted in Fig. 5f (or 5a), then the excess spin up electrons are momentum-locked to flow toward detector Py electrode, giving rise to a low electric charge accumulation at the detector.

From a separate set of measurements of the magnetic lead magneto-resistance, we have determined that the magnetizations of the Py free layers of leads E2 and E3 switch at nearly identical magnetic fields applied parallel to the magnetic lead nanowires. Therefore, antiparallel alignment of the Py magnetizations of the injector and detector leads is not considered for the measurements in Fig. 4 (Co reference leads remain pinned in the in-plane direction perpendicular to the $Sb_2Te_3$ nanowire for all measurements reported in this paper). In non-local spin valves with topologically trivial channel materials such as Cu or graphene, simultaneous reversal of the injector and detector magnetizations does not change the non-local voltage $V_{NL}$.[32, 33, 35] Therefore, our measurement showing a step in $V_{NL}$ upon simultaneous reversal of both Py layers is unusual and reveals the unique characteristic of a spin-momentum locked channel.

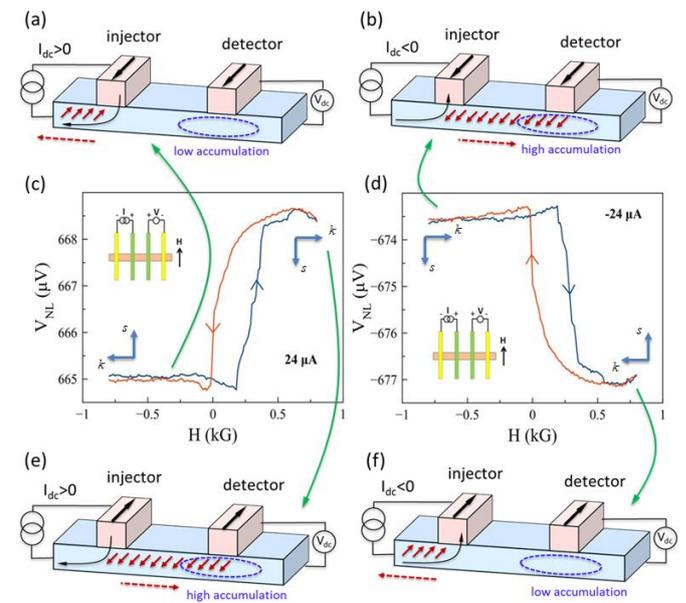

**Fig. 5**. Illustration of the origin of non-local voltage $V_{NL}$ signal in the $Sb_2Te_3$ nanowire. (a) For positive bias $I_{dc}$ > 0 and $M_{Py}$ down, spin down electrons tunnelling out to the Py electrode, leaving excess spin up electrons, with spin $s$ orthogonally locked to the momentum $k$, flowing away from the detector, resulting in a low charge accumulation at the detector, i.e. a lower $V_{NL}$ signal measured in data shown in (c); (b) For negative bias $I_{dc}$ < 0 and $M_{Py}$ down, spin down electrons injected into the channel, momentum-locked to flow toward the detector, giving rise to high charge accumulation at the detector, i.e. a higher $V_{NL}$ signal in (d). Analogously, scenario (e) and (f) respectively depicts the counterpart of (a) and (b) with reversed $M_{Py}$ direction, exhibiting a step change in the $V_{NL}$ signal.

The injector-detector separation in our devices is 0.5 μm, which demonstrates that spin currents in the surface state of the $Sb_2Te_3$ nanowire can flow over long distances. Given the strong spin polarization of such currents, significant spin accumulation can be achieved in the $Sb_2Te_3$ nanowire channel in this non-local geometry. Consequently, the enhanced spin accumulation under the detector electrode gives rise to the observed non-local voltage variation upon simultaneous reversal of the Py magnetization in both ferromagnetic leads.

## Conclusions

This work demonstrates the unique spin-momentum locking property in topologically protected surface states of $Sb_2Te_3$ nanowires by injecting spin-polarized electrons through magnetic tunnel junction electrodes. The observed step increase in non-local voltage $V_{NL}$ signal exhibits a symmetry, which is qualitatively different from that of a non-local voltage measurement in a channel that lacks spin-momentum locking. Simultaneous reversal of magnetic moments of all magnetic contacts to the $Sb_2Te_3$ nanowire alters the non-local voltage, in sharp contrast to non-local signals in conventional non-local spin valves. This unusual symmetry provides a clear signature of the spin-momentum locking in the $Sb_2Te_3$ nanowire topological surface states. Such unique property can be applied as a potential candidate of bit-line channel material in spin-transfer-torque random access memory, as well as for spin-based quantum computation.

## Conflicts of interest

The authors have conflicts to declare.

## Acknowledgements

We would like to thank Dr. Martina Luysberg and Abdur Jalil for high resolution TEM imaging. This work was supported by the National Science Foundation through Grants No. DMR-1610146, No. EFMA1641989 and No. ECCS-1708885. We also acknowledge support by the Army Research Office through Grant No. W911NF-16-1-0472, Defense Threat Reduction Agency through Grant No. HDTRA1-16-1-0025.